# MR-zero meets FLASH – Controlling the transient signal decay in gradient- and rf-spoiled gradient echo sequences

Simon Weinmüller[1], Jonathan Endres[1], Nam Dang[1], Rudolf Stollberger[2], Moritz Zaiss[1,3]

[1]*Institute of Neuroradiology, University Clinic Erlangen, Friedrich-Alexander-Universität Erlangen-Nürnberg (FAU), Germany*

[2]*Institute of Biomedical Imaging, Graz University of Technology, Graz, Austria*

[3]*Department Artificial Intelligence in Biomedical Engineering, Friedrich-Alexander-Universität Erlangen-Nürnberg, Erlangen, Germany*

Simon.weinmueller@uk-erlangen.de

# Abbreviations

- flip angle: $\alpha$
- quadratic phase increment: $\Psi$
- phase angle: $\varphi$
- flip angle train: $\vec{\alpha}$
- phase angle train: $\vec{\varphi}$
- relative B1: rB1
- root mean square error: RMSE

# Abstract

## Purpose

The complex signal decay during the transient FLASH MRI readout can lead to artifacts in magnitude and phase images. We show that target-driven optimization of individual rf flip angles and phases can realize near-ideal signal behavior and mitigate artifacts.

## Methods

The differentiable end-to-end optimization framework MR-zero is used to optimize rf trains of the FLASH sequence. We focus herein on minimizing deviations from the ideally spoiled signal by using a mono-exponential Look-Locker target. We first obtain the transient FLASH signal decay substructure, and then minimize the deviation to the Look-Locker decay by optimizing the individual (i) flip angles, (ii) rf phases and (iii) flip angles and rf phases. Comparison between measurement and simulation are performed using Pulseq in 1D and 2D.

## Results

We could reproduce the complex substructure of the transient FLASH signal decay. All three optimization objectives can bring the real FLASH signal closer to the ideal case, with best results when both flip angles and rf phases are adjusted jointly. This solution outperformed all tested conventional quadratic rf cyclings in terms of (i) matching the Look-Locker target signal, (ii) phase stability, (iii) PSF ideality, (iv) robustness against parameter changes, and (v) magnitude and phase image quality.

Other target functions for the signal could as well be realized, yet, their response is not as general as for the Look-Locker target and need to be optimized for a specific context.

## Conclusion

Individual flip angle and rf phase optimization improves the transient signal decay of FLASH MRI sequences.

# Introduction

The Fast Low-Angle Shot (FLASH) sequence – or gradient- and rf-spoiled gradient echo sequence – is one of the major sequences in clinical applications and was one breakthrough that made fast MRI clinically feasible[1–5]. In steady-state mode, a FLASH sequence provides T1 or susceptibility weighted imaging. In transient-state mode FLASH is especially relevant in magnetization-prepared versions like Magnetization Prepared – Rapid Gradient Echo (MPRAGE) sequence, which provide unique strong gray/white matter contrast[6]. In addition, magnetization transfer contrast or chemical exchange saturation transfer sequences can be realized with prepared transient FLASH sequences[7–10]. Because the prepared initial state provides the contrast, the transient echo trains are limited in their length leading to either small volumes[1,7], or require multi-shot mode to fill larger k-spaces.

A common problem in transient FLASH is that the decay of the magnetization still affects quality of the image, as the decay in k-space acquisition leads to ringing or blurring artifacts, moreover the prepared magnetization state is decaying during acquisition. The decay to the FLASH steady-state signal $S_{SS}$ is described by the Look-Locker decay rate given by the joint effects of T1 and excitation and reads[11,12]:

$$S_{LL}(n \cdot TR) = (S_i - S_{SS}) \cdot \exp\left(-\frac{n \cdot TR}{T_{1,LL}}\right) + S_{SS} \qquad (1)$$

$$\text{with} \qquad \frac{1}{T_{1,LL}} = \frac{1}{T_1} - \frac{\ln(\cos \alpha)}{TR}. \qquad (2)$$

$S_i$ is the initial signal and $n$ is the number of repetition.

Interestingly, this mono-exponential reflects an idealized description in the case of ideal spoiling. In real measurements the signal decay follows coarsely this exponential function, but has complex substructure that depends, among other factors, on the specific rf-cycling[4]. The familiar Look-Locker decay expression is only valid for $T_2 \to 0$[12]. Moreover, this decay is also only observed for constant flip angles, and the decay can effectively be altered by so-called variable flip angle approaches[13,14]. By changing the flip angles and phases of each rf pulse in the train, the intrinsic transient decay substructure can be modified.

In transient FLASH sequences this possibility was mainly used for two objectives: (i) to realize the best approximation of ideal spoiling meaning approaching the Look-Locker mono-exponential decay, or (ii) to realize a specific target signal decay and thereby improved point spread functions (PSF) and image quality. We want to herein revisit objective (i) using the end-to-end framework MR-zero[15].

MR-zero consists of a MR simulation, based on phase distribution graphs[16], which accurately describes the signal behavior of arbitrary MR sequences, allowing both, encoded and unencoded signal analysis. The simulation considers isotropic diffusion, since diffusion can have an influence on the spoiling process[17]. In contrast to extended phase graphs[18–22], which can only describe echo amplitudes, phase distribution graphs can describe the whole echo shape by extending for dephased magnetization and arbitrary timing. This allows to analyze how these signal variations affect the MR image in magnitude and phase. The MR-zero implementation is written in PyTorch[23], which provides full differentiability via auto-differentiation in all input parameters allowing gradient descent optimization of all individual sequence events, e.g. here each individual rf pulse angle and rf pulse phase, when providing a target image or signal. Thus, flexible flip angles and rf phases beyond constant trains and quadratic rf cycling can be investigated by simulating MR images.

Specifically, in this article we want to answer the following question: Can we reproduce the transient decay substructure of the FLASH signal, understand its effect in the image domain of and further improve it using MR-zero?

# Methods

As a first step, this study compares the MR-zero simulation with measurements and results from Epstein et al.[4]. These experiments were performed without phase encoding gradients.

In a second step we use MR-zero to optimize the rf trains using the ideally spoiled unencoded signal as target, and compare the new strategies to existing ones.

In a third step, we investigate how the performance of the different found rf trains translates to 2D imaging in simulation and in vivo. The optimized trains are therefore used in a centric-reordered 2D FLASH readout scheme. As an outlook, we also show an optimization using another target signal, namely, a constant signal target, and a Hanning-shaped signal target in the Supporting Information.

## MR sequence

*Unencoded measurements:*

The sequence parameters as described by Epstein et al.[4] are employed for both simulation and measurement: $\alpha = 10°$, TR = 10 ms, TE = 5 ms, readout bandwidth = 217 Hz/pixel, FOV = 200 mm, slice thickness = 10 mm, matrix = 128 x 128, rf spoiling with a quadratic increment of $\Psi = 117°$, $\Psi = 84°$ and $\Psi = 50°$ and gradient spoiler. We used a saturation preparation with recovery time $T_{REC}$ = 970 ms while Epstein[4] used in his experiments an inversion pulse resulting in similar prepared magnetization. The unencoded signal ($k_x = 0$, $k_y = 0$) is given by the center ADC signal at the echo top where $k_x = 0$ and as all phase encoding gradients were switched off, also $k_y = 0$.

$\Psi = 50°$ rf increment was added since some vendors use and papers recommend it[24–26] and Preibisch et al. suggest it, at least for robust T1 mapping[27]. In order to enable the comparison between measurements and simulations, an additional normalization measurement without preparation was performed with adequate relaxation time, and used for signal normalization purposes. As excitation pulse a slice-selective sinc-shaped pulse with apodization of 0.5 and time bandwidth product of 4 was used. The duration of the pulse was 1 ms.

Repeating the sequence of these unencoded experiments with phase encoding gradients and image reconstruction can be found in the Supporting Information in Figure S1.

*Encoded measurements:*

To make the existing signal fluctuations stronger visible in an image, we decreased TE and increase the flip angle. To achieve shorter TE and TR of TE = 3.2 ms and TR = 6 ms, the readout bandwidth was increased to 500 Hz/pixel. The excitation flip angle was increased to α = 19.5° and no magnetization preparation before image acquisition was used. All other parameters where chosen equally to the unencoded experiment, but phase encoding was switched on and all ADC samples were used in the reconstruction. A centric reordering for acquisition was used. Images with higher resolution of 256 were measured with the same sequence parameter using two shots per image acquisition and the identical rf train for each shot.

Simulation

All described sequences were simulated and also optimized in the MR-zero framework, which consists of a phase distribution graph simulation[16] written in PyTorch[23] which provides full differentiability in all input parameters allowing gradient descent optimization via auto-differentiation. The end-to-end end training procedure was using the ADAM optimizer[28]. The approach extracts additionally useful information about the magnetization evolution like the latent signal evolution within a sequence. Encoded and unencoded simulation can be performed. Using the Pulseq standard[29–31], sequences are directly transferable to a real scanner.

The simulation requires a 3D object defining the MR parameters (proton density, T1, T2, T2', rB1, ΔB0, and the isotropic diffusion constant D) to simulate the MR signal. Two different in silico phantoms are used within this paper, a single voxel phantom and a whole brain phantom. The single voxel phantom is a 3D phantom with one non-zero voxel in the center with one set of MR parameters. This phantom is used for all optimization processes in this script. For this phantom proton density is set to 1 and T2' is 30 ms. The brain phantom is built from the data provided by the BrainWeb database[32]. Segmented maps for grey matter (GM), white matter (WM) and cerebrospinal fluid (CSF) are filled with values for the various physical properties[33–35]. Then they summed together, weighted by their contribution to every voxel as given by the BrainWeb data, resulting in the input maps used by the simulation. $B_0$ (-10.5 Hz – 44.4 Hz) and $rB_1$ (0.84 – 1.14) inhomogeneities were added by using Normal and Cauchy distributions coarsely matching in vivo data.

Experiments / Optimization Objectives

The ideally spoiled signal is for constant flip angles identical to the Look-Locker decay, which we can explicitly generate by the phase distribution graph simulation. In the following $S_{LL}$ is the ideal signal for a constant excitation flip angle α = 10° for the comparison with Epstein and 19.5° for the image quality experiment. For $S_{LL}$ the phase cycling is irrelevant.

Task 1: ($\vec{\alpha} \mid S_{LL}$; $\Psi = 117°$). Optimizing the flip angles $\vec{\alpha}$ only, using $S_{LL}$ as a target, while quadratic phase cycling with $\Psi = 117°$ is preserved. The objective function of this is

$$\vec{\alpha}_{opt} = argmin_{(\vec{\alpha})}(\|S(\vec{\alpha}) - S_{LL}\|) \text{ s.t. } \Psi = 117°. \tag{3}$$

Task 2: ($\vec{\varphi} \mid S_{LL}$; $\alpha = 10°/19.5°$). Optimizing the rf phases $\vec{\varphi}$ only, using $S_{LL}$ as a target, while the flip angle $\alpha = 10°/19.5°$ is preserved. The objective function of this is

$$\vec{\varphi}_{opt} = argmin_{(\vec{\varphi})}(\|S(\vec{\varphi}) - S_{LL}\|) \text{ s.t. } \alpha = 10°/19.5°. \tag{4}$$

Task 3: ($\vec{\alpha}, \vec{\varphi} \mid S_{LL}$). Optimizing both flip angles $\vec{\alpha}$ and rf phases $\vec{\varphi}$, using $S_{LL}$ as a target. The objective function of this is

$$\vec{\alpha}_{opt}, \vec{\varphi}_{opt} = argmin_{(\vec{\alpha},\vec{\varphi})}(\|S(\vec{\alpha}, \vec{\varphi}) - S_{LL}\|). \tag{5}$$

Thus, each task aims for the smallest deviation from the ideally spoiled signal $S_{LL}$. All other parameters of the sequence such as timing, bandwidth, etc. are kept fixed, thus if $\vec{\varphi}$ is not optimized the quadratically growing rf phases of each pulse is used with $\Psi = 117°$.

As loss function $\perp (\vec{x}, \vec{y})$ we used the so-called perpendicular loss proposed for complex data by Terpstra et al.[36]. It shows a symmetric loss landscape, achieving better performance and faster convergence. This loss is defined as

$$\perp (\vec{x}, \vec{y}) = P(\vec{x}, \vec{y}) + \ell^1(|\vec{x}|, |\vec{y}|), \tag{6}$$

$$P(\vec{x}, \vec{y}) = \frac{|\text{Re}(x)\text{Im}(y) - \text{Im}(x)\text{Re}(y)|}{|\vec{y}|}. \tag{7}$$

## Measurement

All measurements were performed at a 3T MAGNETOM PRIMSA scanner (Siemens Healthcare, Erlangen) using a 20-channel Rx head coil. The used scanner shows a significant B0-drift from one measurement to another. In order to obtain a match between measurement and simulation, the simulated phase was shifted by the amount of the first measured value so that the first values in simulation and measurement matches. This shift differs from one measurement to another.

For in vitro measurements an agar-based phantom was created consisting of a 50 ml falcon tube. 13.5 mg Copper sulfate was used to control T1 and T2 relaxation value while 281.1 mg agar was added to avoid convection. Measurements of this tube phantom were compared with unencoded simulation where the tube is approximated by a digital phantom consisting of one with phantom parameter filled voxel. To facilitate a comprehensive comparison between measurement and simulation, phantom parameters were gained by a signal fit of quantitative measurements (T1 $\approx$ 1.0 s, T2 $\approx$ 0.17 s, D $\approx$ 1.6 $\cdot$ 10$^{-3}$ mm$^2$/s). The slice selective pulse efficiency was measured with a standard actual flip angle method resulting in a relative B1 value of approximately 0.77, which was used to match simulation and experiment flip angles.

Three healthy subjects (m, 23; m, 31 and w, 23) were scanned within this paper after written informed consent and approved by the local ethics committee. In vivo measurements correspond accordingly to the simulation with the data generated with BrainWeb[32]. For the optimization task, again just one representative voxel was used with parameters matching the median of the in brain phantom (T1 ≈ 1.5 s, T2 ≈ 0.11 s, D ≈ $0.84 \cdot 10^{-3}$ mm²/s).

## Results

In a first step, we obtain the finding of Epstein et al.[4], who first demonstrated that even with gradient- and rf-spoiling the transient signal decay deviates from an ideally spoiled exponential Look-Locker decay. In Figure 1, we show that we find similar deviations from the ideal Look-Locker decay as Epstein et al. both experimentally, as well as using the phase distribution graph simulation of MR-zero. The high agreement between the experimental data and simulation let us conclude that our simulation framework is able to describe the FLASH signal realistically and generalization to variable flip angles and altered rf cycling can be expected.

In the next step, we want to use the MR-zero framework to optimize individual flip angles and rf phases so that the transient FLASH decay matches the ideal signal $S_{LL}$. As described above we tested three different objectives. Task 1 optimize only the flip angles $\vec{\alpha}$, for a quadratic phase cycling with increment $\Psi = 117°$. Task 2 optimizes each individual phase of rf phase angle train $\vec{\varphi}$, meaning we explicitly leave the condition of quadratically increasing phases. Task 3 optimizes both ($\vec{\alpha},\vec{\varphi}$).

Figure 2 depicts the evaluations, compared to the ideal mono exponential decay $S_{LL}$ (column 2) and to a constant phase (column 3). The first column shows the optimized flip angle and rf phase trains for all three tasks. The overall result of Figure 2 is that for all tasks we get closer to the target signal upon optimization. Task 1 ($\vec{\alpha}$) leads to signal close to the target, but no constant phase. Task 2 ($\vec{\varphi}$) leads to a constant phase while improving the signal with a RMSE similar to a rf cycling of $\Psi = 84°$. Task 3 ($\vec{\alpha},\vec{\varphi}$) leads to the best results regarding target signal and constant phase. The last task was initialized with the phases found in the second task, thus similar phases and only small flip angle variations are observed here. This can also be seen quantitatively in Table 1A, where the RMSE of measurement and simulation was calculated.

The PSFs of the different rf increments and optimization tasks, including the difference to the ideal PSF are shown in Figure 3. The PSFs of typical quadratic phase cycling approaches show deviations from the ideal PSF. Of the three tested $\Psi$, a rf cycling increment of $\Psi = 84°$ performs best with regard to RMSE. Yet, the optimization tasks can further improve the signal: especially with phase optimization we get closer to an ideal PSF upon optimization. Task 1 slightly improves the PSF. However, the non-constant signal phase leads to deviations in the PSF. In contrast, task 2 and 3 improve signal phase and PSF. Overall, task 3 leads to the best results regarding target signal, constant phase and therefore also PSF.

The found strategies in task 2 and 3 outperform typical quadratic phase cycling approaches, which underlines that an individual pulse optimization approach can be beneficial even for FLASH.

Before we test the stability of the found solutions against system changes, let us first investigate the underlying strategies. For this, the phase graph analysis provided by the $\tau$-dephasing *latent signal* plot of phase distribution graphs is insightful (Figure 4)[16]. The latent signal provides insight, which echoes can contribute to the signal in the current or any later repetition. For a gradient-spoiled FLASH, this must be rephased echoes restored back into the z-magnetization. The ideal FLASH/Look-Locker signal, consisting only of fresh FIDs, would correspond to only one line at $\tau = 0$. Thus, every intensity at $\tau \neq 0$ reflects unwanted signals that can interfere with each other, especially with the FID signal. These originate from higher and higher echoes given by $\tau$, which indicates the dephasing time of transverse magnetization.

Already small differences can be observed in the latent signal plots for different quadratic rf spoiling of 117°, 84° and 50° in Figure 4a-c, with 50° and 84° appearing smoother along the repetition dimension. The latent signal plots of the optimized sequence are shown Figure 4d-f. The flip angle optimization task 1 doesn't change the structure of the diagram, thus the higher echo contribution is similar – most probably the large FID signal is modulated by the flip angle change to counteract restored z-magnetization. For the latter two tasks including rf spoiling optimization, the latent signal plots change quite significantly. More and stronger dephased paths contribute to the final signal. Thus, the underlying strategy is most probably a tailored destructive and constructive interference of more different signals, which leads to the target signal.

Let us now analyze the stability of the observed behavior when changing system parameters, namely, the relaxation times T1 and T2, the initial magnetization before the readout MI, and the B1 inhomogeneity rB1. The conventional quadratic rf cycling approaches show no surprises against these system changes (Figure 5). In principle the same deviations as above are observed, only their magnitude is affected, mostly by T2 and rB1. As previously, the $\Psi = 117°$ shows the strongest deviations with regard to the exponential decay of $S_{LL}$.

The stability against system parameter changes is more diverse for the optimized FLASH signal (Figure 6). Most problematic is the behavior for task 1 ($\vec{\alpha} \mid S_{LL} ; \Psi = 117°$); the signal patterns appear smooth, but deviations grow when leaving the system parameters of the optimization. More severely, the signal modulation appears inverted for T1, T2, MI, or rB1 below or above the initial value (yellow line in all plots). This fits to our previous interpretation that the flip angle alters the FID signal, to counteract the restored signals. But, if the restored signal composition now changes due to the system parameters, this compensation is unbalanced and turns out to be not robust.

Most interestingly, task 2 ($\vec{\varphi} \mid S_{LL}$ ; $\alpha = 10°$) and task 3 ($\vec{\alpha}, \vec{\varphi} \mid S_{LL}$) show also here the best performance and robustness. Similar as for conventional rf cyclings also these optimized cases behave even more robust against the introduced changes and lead to signal and phase most similar to the ideal LL signal in all cases. Interestingly, task 3 outperforms the pure rf phase optimization (especially visible in the more smooth signal magnitude in Fig 6A(d,h,l,p), despite the fluctuations observed for the pure FA optimization. However, this is mostly due to the much smaller flip angle variation in task 3 compared to task 1.

To summarize our results: All three objectives can bring the real FLASH signal closer to the ideal case, but pure flip angle approaches are less robust against parameter changes. Best results are achieved when first the phase cycling is optimized and then the flip angle pattern is slightly adjusted jointly with the phases. This solution also outperformed all tested conventional rf cyclings in terms of (i) matching the target signal, (ii) phase stability, (iii) PSF ideality, and (iv) robustness against parameter changes, as quantified by the average RMSE given in Table 1B.

In a last step we investigate the influence of the different rf trains on image quality. For the chosen sequence parameters TE = 5.0 ms, TR = 10 ms, bandwidth of 500 Hz/pixel, and FA $\alpha$ = 10.0° the effect is less visible. Susceptibility and B0 inhomogeneity effects are in the same order (data shown in Supporting Information in Figure S1). Therefore, an adaption of the sequence was performed by reducing TE and TR and increasing the bandwidth and FA $\alpha$. The optimization was performed for the same three tasks as in the first part, but employing a voxel phantom with median phantom parameters (T1 $\approx$ 1.5 s, T2 $\approx$ 0.11 s, D $\approx$ 0.84 $\cdot$ $10^{-3}$mm$^2$/s). In this case, measurements were taken for both, the unencoded sequence and a centric reordered encoded sequence. The FLASH magnitude and phase images can be seen in Figure 7 for simulation (Figure 7A, B) and measurement (Figure 7C, D). The rf increment of 117° produces clear artefacts in the magnitude image, as well as in the phase image, clearly visible in the difference images. This is reduced especially by task 2 and 3, but not for task 1. This is in line with the 1D result that flip angle optimizations are not robust against system parameter changes.

While some of the artifacts look like motion artifacts, motion between the images was not visible during the 3 minutes long measurement. Furthermore, similar artefacts are visible in the simulation where no motion is considered.

For the sake of completeness, we also provide the unencoded signals, reflecting the average signal values for the whole slice of the in vivo measurement in Figure 8. Figure 8 is in line with previous results in simulation and in vitro: flip angle optimization alone fails when system parameters are changed, task 2 and 3 lead to improved signal decays. $\Psi = 84°$ and $\Psi = 50°$ looks most smooth here, while task 2 and 3 introduce noise-like patterns in higher repetitions, resembling changes towards shorter T2 (see Figures

6Ad and Bd). This noise like fluctuation seems to have no clear influence on the images in Figure 7g). Task 2 and task 3 improves especially the smoothness of the decay along the first repetitions. There, a very smooth signal profile for task 3 is visible (Figure 8k). In comparison to that, an increment of 117° exhibits a pronounced edge in the unencoded signal and significant phase variations of up to 20°.

In summary, 84° delivers a good solution for a smooth transient decay, as already suggested by Epstein et al.[4]. By the optimization of task 2 and especially task 3, in vivo an improvement of the signal decay can be achieved, even compared to 84°. The non-generalization of task 1 is clearly visible in the signal and images.

In Figure 9 the impact of image quality for a high-resolution image of 256 by using a 2-shot sequence resulting in less blurry images was investigated. The strength of the signal phase artefacts in the unencoded experiment corresponds to the artefacts observed in the centric encoded FLASH phase images. As the phase image has less contrast, such artifacts are easiest to spot in the phase image. An ideal Look-Locker experiment should have a constant phase in the unencoded case, thus the smoothest phase image. Phase artefacts translate to artefacts in the magnitude image. The phase image of $\Psi = 117°$ shows artifacts in both unencoded phase (Figure 9Aa) and image phase (Figure 9Ba), which is improved by $\Psi = 84°$ (Figure 9Ab and 9Bb), and outperformed by our optimized sequence (Figure 9Ac-d and Bc-d). Remaining phase artefacts of a quadratic rf increment of 84° vanish for the optimization task as can be seen in the Figure 9Ca and 9Cd where example profile line plots of magnitude and phase images are shown.

Finally, instead of the Look-Locker decay we can also use other target functions, e.g. a constant signal target or Hanning function target[37,38]. For this we perform an unencoded optimization with no magnetization preparation of another task 4 ($\vec{\alpha} \mid S_{const}$ ; $\Psi = 84°$) and task 5 ($\vec{\alpha} \mid S_{Hanning}$ ; $\Psi = 84°$) for a voxel filled with tissue parameters of the median of a brain phantom (GM-like T1 value). As seen in Supporting Information Figure S2, this goal can be reached with a flip angle optimization, but just for the tissue parameter of the training. The solution is tissue- and problem-dependent and leads to non-constant signals for other T1 values (CSF and WM). Thus, this optimization has to be performed within specific contexts, which is beyond the scope of this work. The Look-Locker target leads to generally applicable solutions.

## Discussion

### Overview and Interpretation of Findings

In this article we revisited the problem of the transient FLASH signal decay using the end-to-end MR-zero framework with a phase distribution graph simulation. The advantage of methodology by using phase distribution graphs instead of an isochromat or spin simulation, allows us to achieve a higher level of precision in our simulations compared to Loktyushin[15], as shown in the Supporting Information in Figure S3. To describe proper spoiling, a high number of spins is necessary, which requires significant simulation time. In comparison, the PDG simulation is more efficient. The accuracy of our simulation is validated within this manuscript by the high agreement of simulation and measurement, even for substructures arising from interference of higher echoes. We showed that a general target-driven signal optimization is possible by optimization of each and every rf pulse angle and phase of the pulse sequence. We further demonstrate in simulation and measurement how artifacts from unencoded experiments transfer to the image domain and differs between optimized rf pulse trains and typical rf cyclings. Despite the general signal optimization potential, we focused herein on the Look-Locker target as discussed in the following.

As an MRI sequence the FLASH or gradient- and rf spoiled gradient echo sequence can be seen as one of the simplest sequences attempting to use only the FID signal of fresh z-magnetization, and eliminating the signal contribution of all higher echoes. However, in reality higher echoes play a role as some of them can get rephased and restored into the non-encoded z-magnetization, which is the origin of the FID signal. In this article, we showed that the MR-zero simulation based on phase distribution graphs[16] is able to accurately simulate these additional signals leading to deviations from the exponential Look-Locker decay in agreement with the findings of Epstein et al.[4]. To achieve the same agreement also for larger flip angles and/or longer TR, isotropic diffusion is included in the simulation since for these parameter regimes a diffusion dependency is expected[17]. Moreover, the differentiability of our simulation allows to alter these deviations via a gradient descent optimization of the individual rf flip angles and phases. From our results we can conclude that such an individual optimization can further improve the signal and image quality of FLASH sequences.

Interestingly, our rf phase cycling optimization challenges the often-used concept of quadratic phase cycling. Thus, we want to point out that quadratic phase cycling was actually derived to generate a signal independent of the repetition *n*, thus, it is derived for the *steady state signal* [39]. Quadratic phase cycling is still used very generally, also in transient sequences, despite the lack of theoretical foundation in this case. Moreover, the exact value of $\Psi$, most famous $\Psi = 117°$, is also derived using a steady-state argumentation, by finding the signal that is most close to the dynamic FLASH equation or Look-Locker equation[3]. Despite the work of Epstein et al.[4] who already suggested the $\Psi = 84°$ phase increment to be better in transient cases, in many transient state applications, even in vendor implementations of magnetization prepared sequences, still quadratic phase increments of $\Psi = 117°$, $\Psi = 50°$ or $\Psi = $

123.5° are used[3,24–26,40]. We conclude, neither certain values of Ψ have a fundamental justification for the transient phase of FLASH signals, nor the quadratic phase cycling itself. Both are derived for steady-state. Thus, the optimization we performed regarding the phase cycling $\vec{\varphi}$ in transient state was actually not yet covered by the literature. Only Epstein et al.[4] tested different Ψ also in the transient case.

An open question might be here how the transition to a steady-state with fixed Ψ is handled, however, this could be included in a task, i.e. to optimize only the first 200 events of an otherwise quadratic phase cycling of a fixed Ψ. As shown in Supporting Information Figure S4 this might even not be necessary, as for typical transient sequences with low flip angle in vivo, the steady-state is almost independent of Ψ[17]. Strong variations of the steady state for different increments are just visible for high flip angles or high TR/T1-ratio[5].

Related Work

In this article we focused on using the ideal Look-Locker[11] decay as a target function, however we also showed in the Supporting Information in Figure S2 that different signal target functions are possible. The general optimization of rf flip angles for this objective was already done by several groups early as 1992 by Mugler et al. for SSFP[13], 1992 by Stehling for rf cycled FLASH[14]. Both Stehling, as well as Priatna and Paschal in 1994[41] show similar flip angle trains as visualize in the Supporting Information in Figure S2. Stöcker and Shah[42] included also the k-space trajectory into their considerations for SSFP and used similar to us also an EPG-based simulation. Li et al.[43] did the same for the FLASH sequence. All the above works aimed for a flat signal response, which leads to a tissue dependent result: flat response in the target tissue leads to a non-flat signal evolution tissues with different relaxation or B1 as discussed by Worters and Hargreaves[44] and shown here in the Supporting Information in Figure S2a-c. Interestingly, none of the above articles optimized the rf cycling, only Epstein et al.[4] identified the best quadratic phase cycling increment Ψ = 84° for the transient decay. While different signal targets have the benefit of PSF-tuning, the Look-Locker target has the benefit of an unchanged Look-Locker decay in all tissues, which is why we focused on this target herein.

This has the benefit, that still all voxel signals are interpretable by the Look-Locker equation (1) , which would be especially relevant for model-based reconstructions assuming the ideal spoiled Look-Locker decay[11,45–47]. For such approaches our refined rf trains might lead to more accurate quantification. Moreover for already existing data with e.g. Ψ = 117°, our differentiable MR-zero simulation could be used directly as improved recon model including the substructure, similar as for Bloch-model-based reconstructions[48].

The present work uses MR-zero on unencoded signals, which is different from previous applications of MR-zero, which aimed for end-to-end target-driven optimization in the image space, for learning encoding, T1 mapping[15], or sharp TSE MRI[49].

### Limitations and Future Directions

Remaining artefacts in rf-spoiled images can be mitigated by either larger gradient spoiling[25], or variable gradient spoiling[50]. Both require higher gradient amplitudes thus the minimal TR increases, which limits fast imaging applications. Controlling the deviation by the rf events therefore allows lower gradient spoiling and faster sequences.

An incidental finding for the transient response was that for zero starting magnetization MI = 0 the signals generated with typical quadratic rf increments or from task 2 and 3 were already most similar to $S_{LL}$ (red curve in Figure 5A(i-l) and Figure 6A(i,k,l)), only task 1 deteriorates the progression (Figure 6A(j)). This is plausible as the fluctuations result from restored echoes of early excitations. If now the early excitations have very small magnitude, the fluctuations vanish. This is interesting as all saturation recovery sequences (e.g. SASHA) fulfill such an initial condition and implicitly should have more accurate exponential recovery curves[51]. Furthermore, for a typical MPRAGE sequence, only slight differences between a quadratic rf increment of 117° and 84° could be observed (data shown in Supporting Information in Figure S5). Since CSF has low signal, but produces most artifacts, only minor influences of different rf spoiling are expected in MPRAGE sequences. However, in tumor or stroke affected areas there might be liquids with different relaxation times and nulling points that are not fully suppressed, but still show the mentioned artifacts.

In contrast, the shown artefact in the magnitude image can lead to errors using a simple Look-Locker model for reconstruction for a series of T1-weighted images. The Look-Locker model leads to T1 map differences for the rf cycling of 117°, which are mitigated when using the data acquired with optimized rf trains. This is shown in the Supporting Information in Figure S6. Nevertheless, there are many more details to be explored in silico and in vivo not only using optimized sequences for simple models, but also using the PDG simulation as reconstruction operator for non-optimized sequences.

Some rf increments lead to visible artifacts in the phase images (Figure 8 and Figure 9). These artefacts can influence results in phase-based sequences as quantitative susceptibility mapping[52] or fast B0 mapping approaches. Our insights might help to improve fast implementations of theses sequences.

## Conclusion

It was shown that the transient signal decay in FLASH sequences shows deviations from the Lock-Locker decay with a complex substructure. These structures are visible in measurement and simulation and are distinct for different rf cyclings. In the past quadratic rf cycling was shown to be able to decrease this substructure, but only for distinct phase increments such as e.g. 84°. Typical phase cyclings even in use in clinical transient FLASH sequences (117° or 50°) show pronounced deviations leading to sidebands in the PSF and image artifacts. By introducing the end-to-end optimization MR-zero we optimize rf phase and flip angles simultaneously leading to a close match to a perfectly spoiled Lock-Locker decay and better PSF resulting in artifact free magnitude and phase images.

## Data Availability Statement

The MR-zero simulation is accessible via https://pypi.org/project/mrzerocore/ and different script can be directly run on a playground Playground MR0.

Optimization results are published here: https://github.com/MRsources/FLASHzero.

## Acknowledgements

Funding by the German Research Foundation (DFG) is gratefully acknowledged (project 500888779 / RU5534 MR biosignatures at UHF, 442377885).

List of Figure:

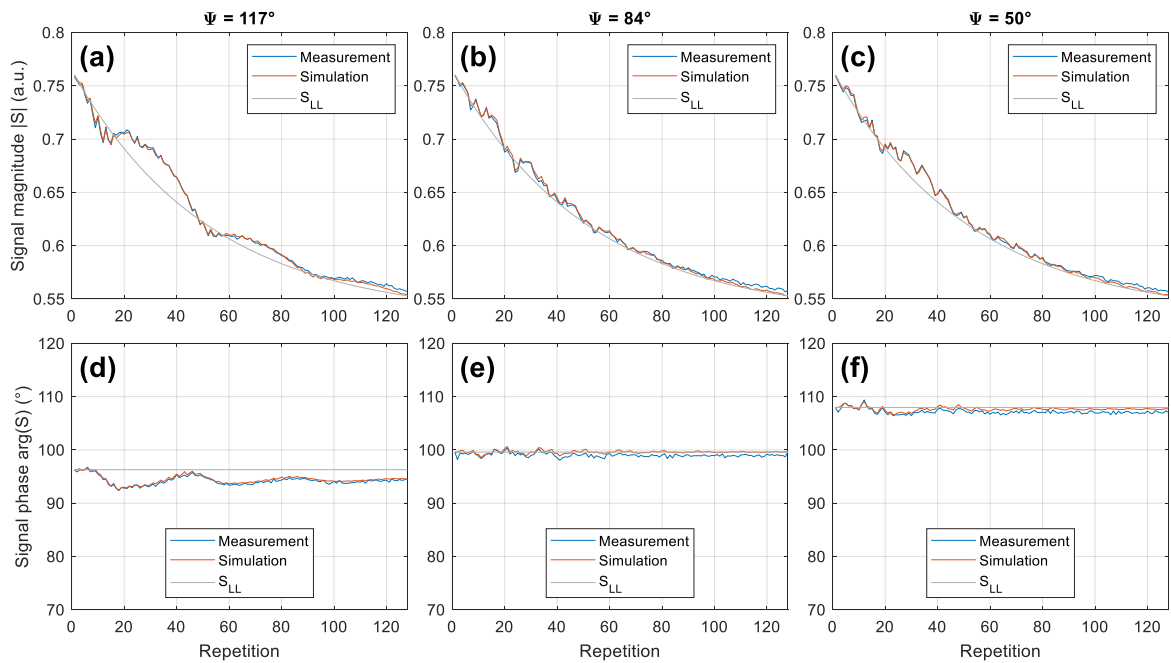

**Figure 1:** Unencoded FLASH signal for α = 10°, TE = 5 ms, TR = 10 ms, after one 90° preparation pulse with a recovery time of 970 ms. Measured and simulated signal magnitude **(a-c)** and signal phase **(d-f)** for Ψ = 117° **(a, d)**, Ψ = 84° **(b, e)** and Ψ = 50° **(c, f)**. Additionally, the ideally spoiled signal $S_{LL}$ is plotted (grey solid line). All results show a good agreement between measurement and simulation, and the respective plots also agree with the results of Epstein et al.[4], where similar deviations from the ideal Look-Locker decay were found. T1 = 0.987 s and T2 = 0.167 s were used in simulation.

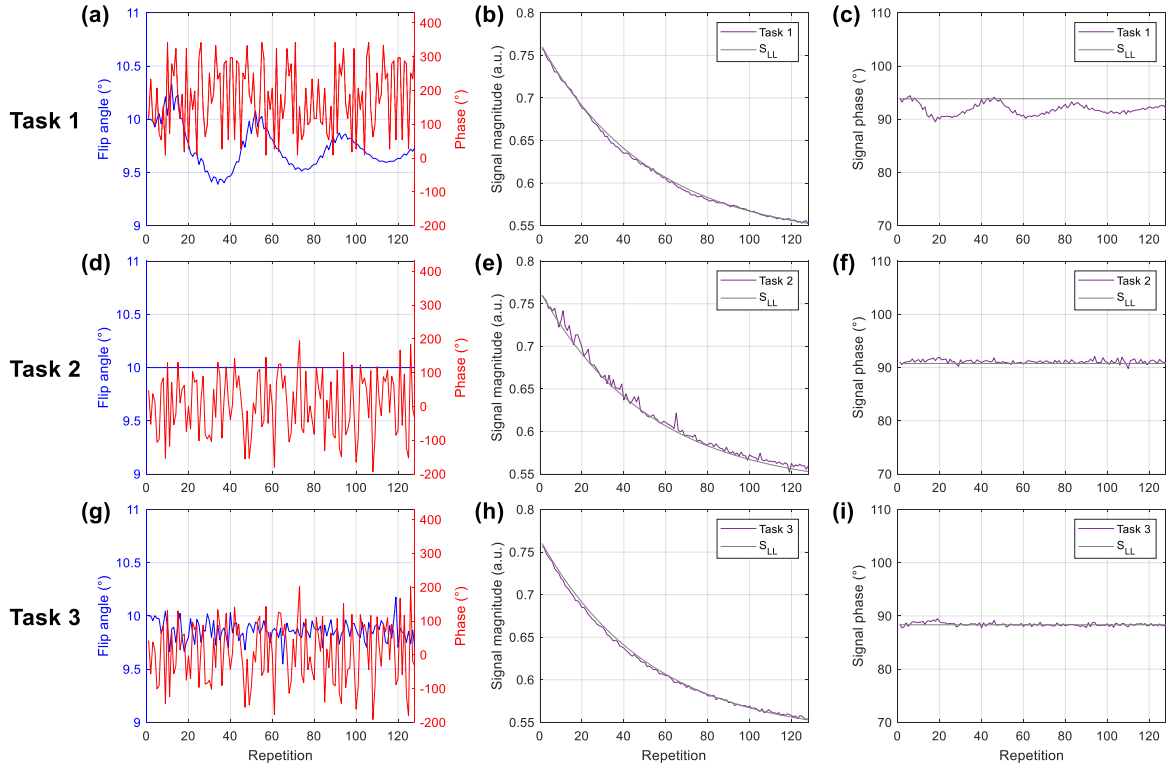

**Figure 2:** The flip angles $\vec{\alpha}$ and rf phases $\vec{\varphi}$ are shown for the three optimization cases **(a)** task 1 ($\vec{\alpha}$), **(d)** task 2 ($\vec{\varphi}$), and **(g)** task 3 ($\vec{\alpha}, \vec{\varphi}$). The measured signal magnitude evaluations are shown in **(b)**, **(e)** and **(h)**, respectively for the three tasks. Additionally, the ideally spoiled signal $S_{LL}$ is plotted (grey solid line). The same comparison is done in **(c)**, **(f)** and **(h)** for the signal phase. While the flip angle optimization adjusts only the signal amplitude, additional rf phase optimization can also alter the signal phase. The best results are achieved with a combined approach, see Table 1B.

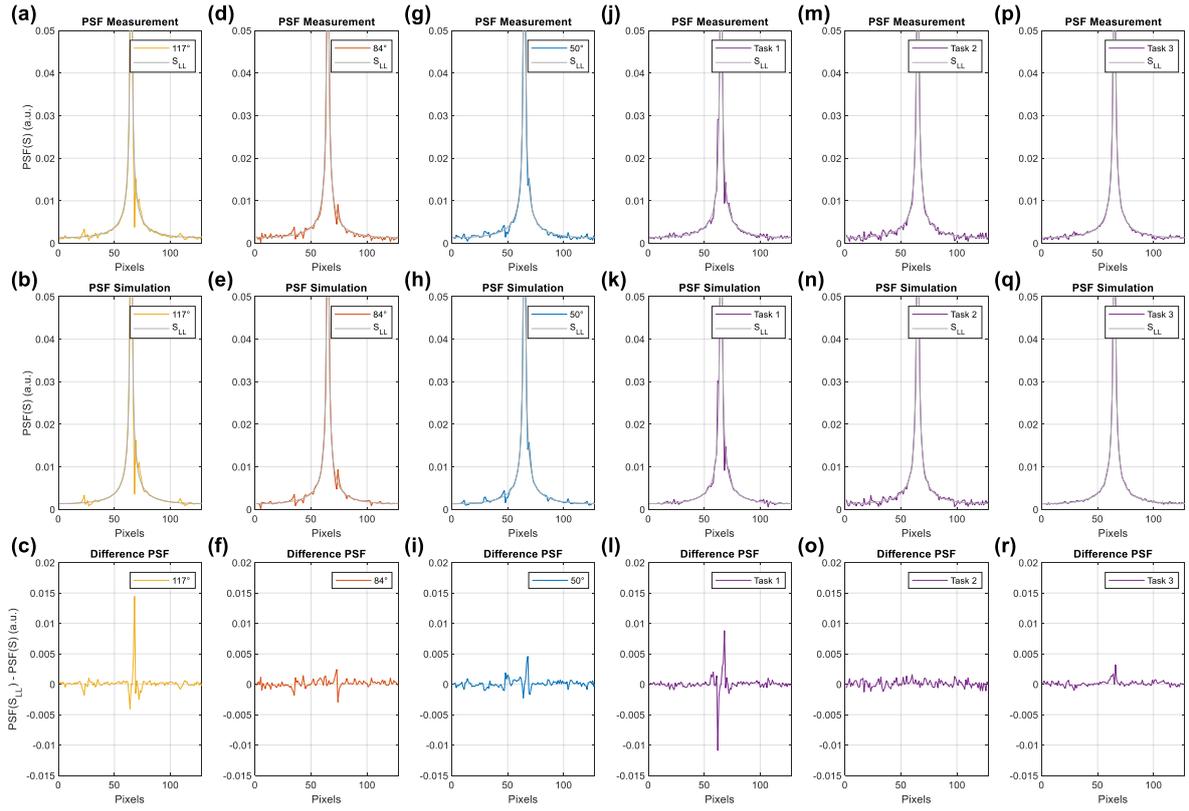

**Figure 3:** Magnitude of PSFs for typical quadratic phase cycling of $\Psi = 117°$ **(a-c)**, 84° **(d-f)** and 50° **(g-i)**. The PSFs of the optimization tasks are visualized in **(j-l)** for task 1, in **(m-o)** for task 2 and in **(p-r)** for the last optimization task 3. The measurement data is shown in the top row, the simulation data in the middle row and the difference between an ideal PSF (grey solid line) to the measured PSF is shown in the bottom row. A closer agreement to the PSF of an ideally spoiled FID can be seen for the optimization tasks, especially when the phases of the pulses are optimized. The best result regarding PSF are achieved by optimization task 3, according to the quantitative results in Table 1A.

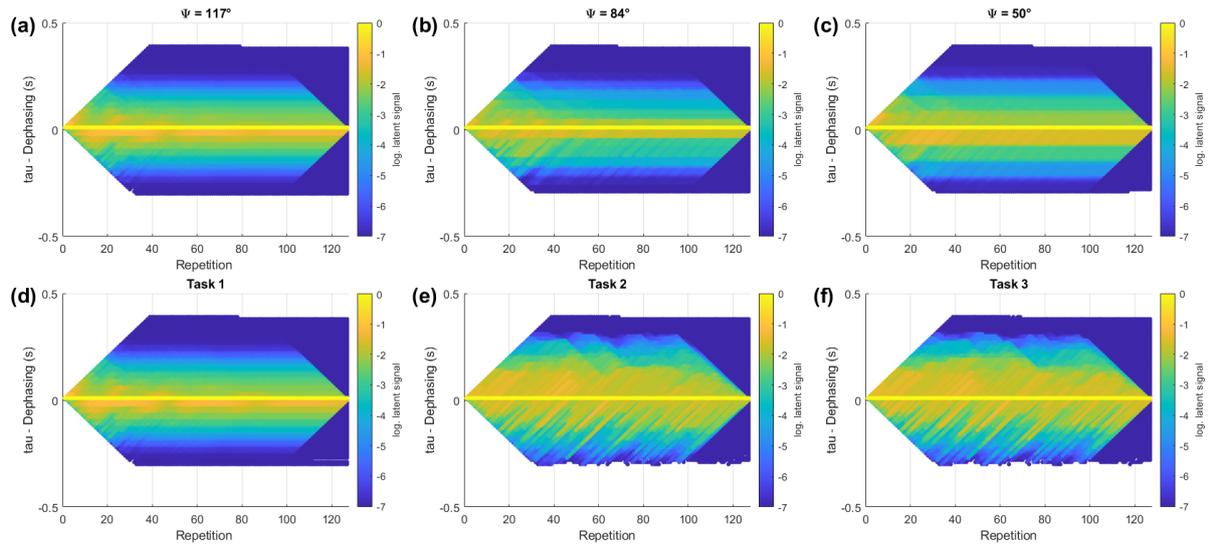

**Figure 4:** Latent signal for FLASH for different rf pulse trains. **(a)** rf increment of Ψ = 117°, **(b)** Ψ = 84° **(b)** and **(c)** Ψ = 50°. Optimization task 1-3 are shown in **(d,e,f)**, respectively. Signals below $10^{-6}$ are not considered, since their impact to the final signal is negligible. The flip angle optimization task 1 does not affect the structure of the diagram while task 2 and 3 do. More and higher dephased and restored echoes contribute to the signal and are the origin of altered signal magnitude and phase.

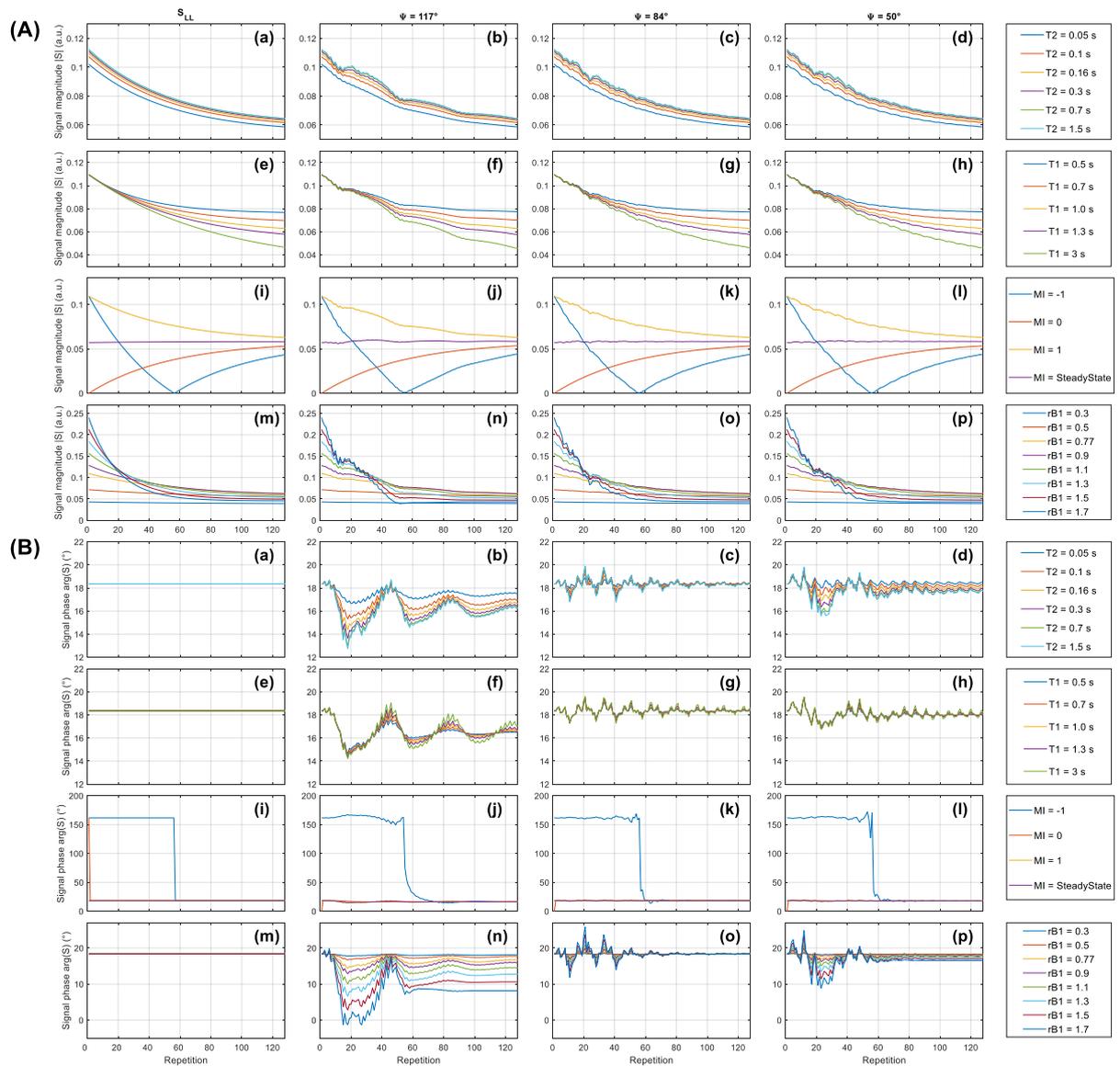

**Figure 5:** Simulation for different T2 values **(a-d)**, T1 values **(e-h)**, initial magnetization values **(i-l)** and relative B1 values **(m-p)** of the signal magnitude **(A)** and signal phase **(B)** for $S_{LL}$, representing a perfect spoiled sequence, and rf cycling of 117°, 84° and 50°. Only one parameter was varied at a time, while the other parameters were kept constant. The yellow curve represents the same curve for the parameters of the tube phantom (T2 ≈ 0.16 s, T1 ≈ 1.0 s, MI = 1, rB1 ≈ 0.77) in all plots. RMSE between system parameters and perfect spoiled FID sequence are shown in Table 1B.

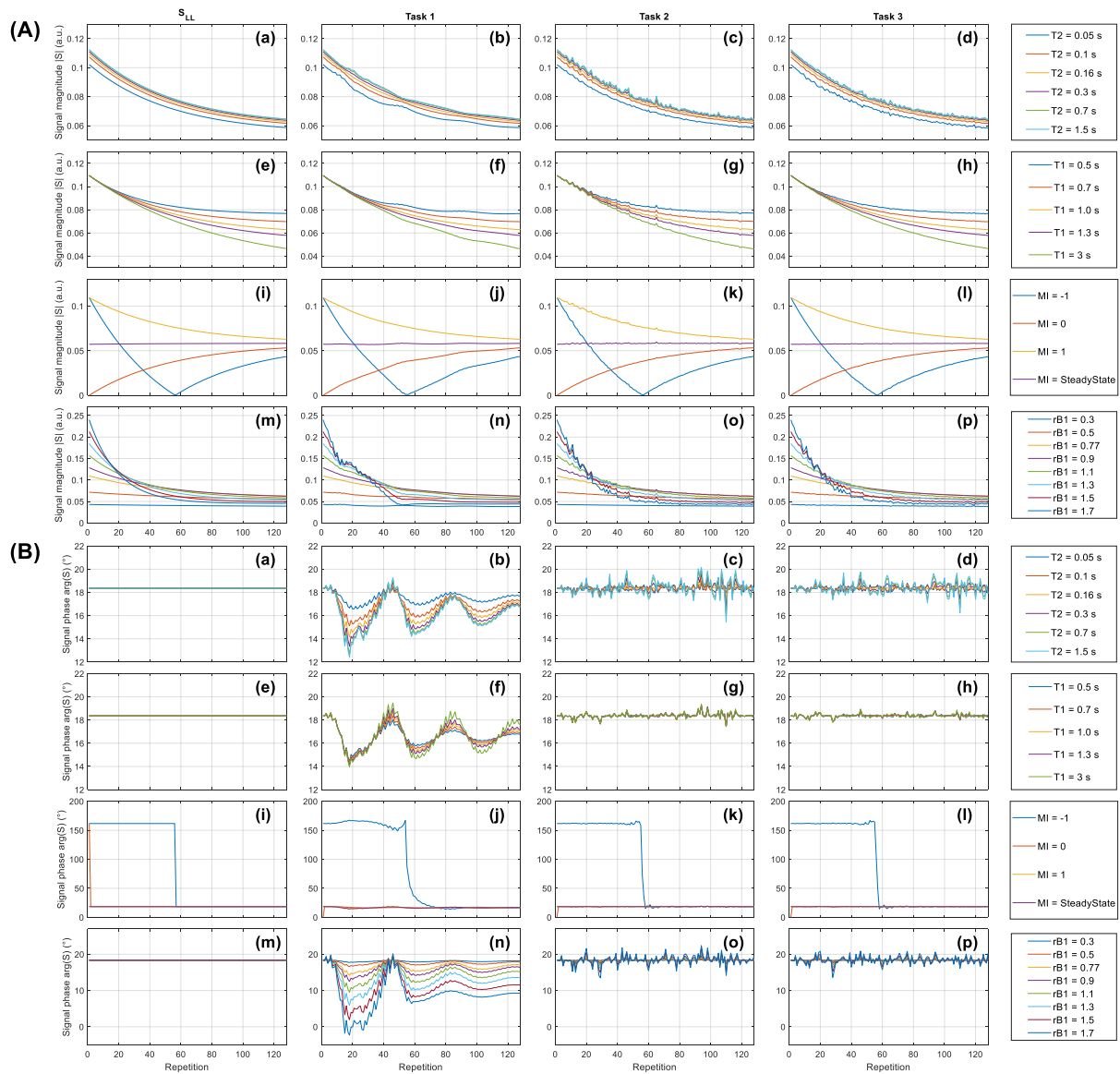

**Figure 6:** Simulation for different T2 values **(a-d)**, T1 values **(e-h)**, initial magnetization values **(i-l)** and relative B1 values **(m-p)** of the signal magnitude **(A)** and signal phase **(B)** for $S_{LL}$ signal, representing a perfect spoiled sequence, and optimization tasks 1-3. Only one parameter was varied at a time, while the other parameters were kept constant. The yellow curve represents the same curve for the parameters of the tube phantom (T2 ≈ 0.16 s, T1≈1.0s, MI = 1, rB1 ≈ 0.77) in all plots. RMSE between system parameters and perfect spoiled FID sequence are shown in Table 1B.

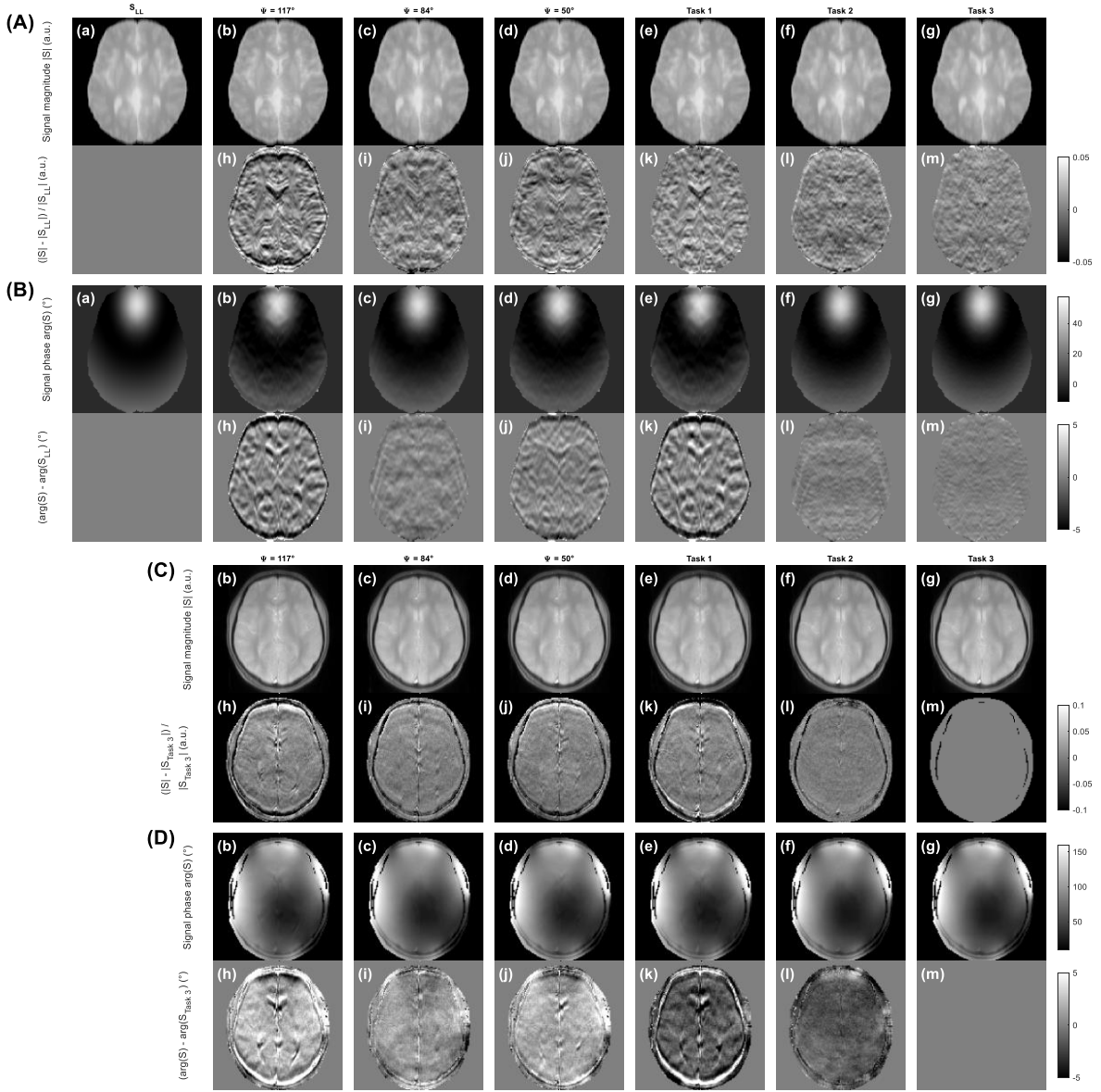

**Figure 7:** Simulation **(A, B)** and measurement **(C, D)** of centric encoded FLASH readouts using α = 19.5°, TE = 3.2 ms, TR = 6 ms for typical rf increments of 117° **(b)**, 84° **(c)** and 50° **(d)** and optimization task 1 **(e)**, task 2 **(f)** and task 3 **(g)**. In simulation the ideal spoiled Look-Locker image $S_{LL}$ was additionally calculated. The phantom for simulation was built from data provided by the BrainWeb database. There, differences of signal magnitude S **(Ah-m)** and phase arg(S) **(Bh-m)** are calculated for each image with respect to the Look-Locker target **(a)**. For the magnitude images **(Cb-g)** and phase images **(Db-g)** task 3 is used as reference **(g)** in measurement since it shows the smoothest phase image indicating the best spoiling. This is also validated by the smoothest phase image in the simulation. An overall good agreement between simulation and measurement can be seen. Artefacts in the phase image can be seen for all standard rf increments, especially for 117°. Task 3 can reduce these artefacts coming from the rf cycling and generate the smoothest phase image. In the Supporting Information in Figure S1 the images corresponding to the unencoded experiment show similar deviations, but less intense especially in phase images due to overlaying susceptibilities and B0 inhomogeneities and longer TE.

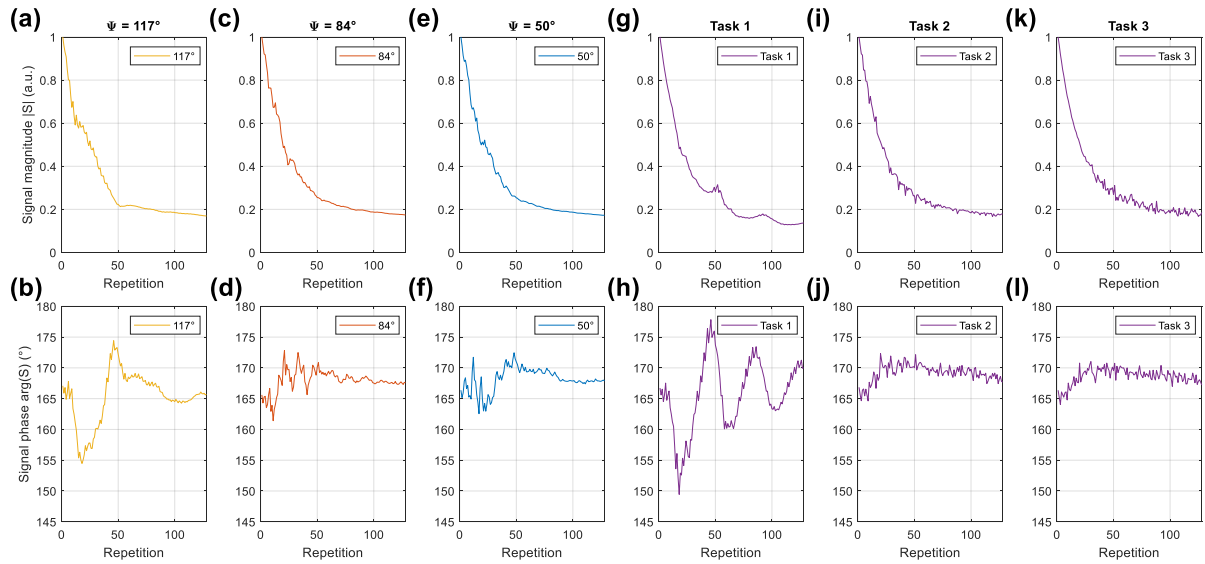

**Figure 8:** Unencoded measurements of the FLASH signal using α = 19.5°, TE = 3.2 ms, TR = 6 ms for typical rf increments of 117° **(a,b)**, 84° **(c,d)** and 50° **(e,f)** and for task 1 **(g,h)**, task 2 **(i,j)** and task 3 **(k,l)** are shown. In the first row, the signal magnitude |S| and in the second row the signal phase arg(S) of each FLASH sequence is visualize. Strong deviations are again visible for 117° increment, while 84° and 50° leads to less variations. Task 1 performs worse w.r.t. a close signal for a Look-Locker decay and constant phase signal. Task 2 and 3 could improve the decay while simultaneously having the most constant phase. For outer k-space lines it results in a more noise like fluctuation which are not visible in images as can be seen in Figure 7.

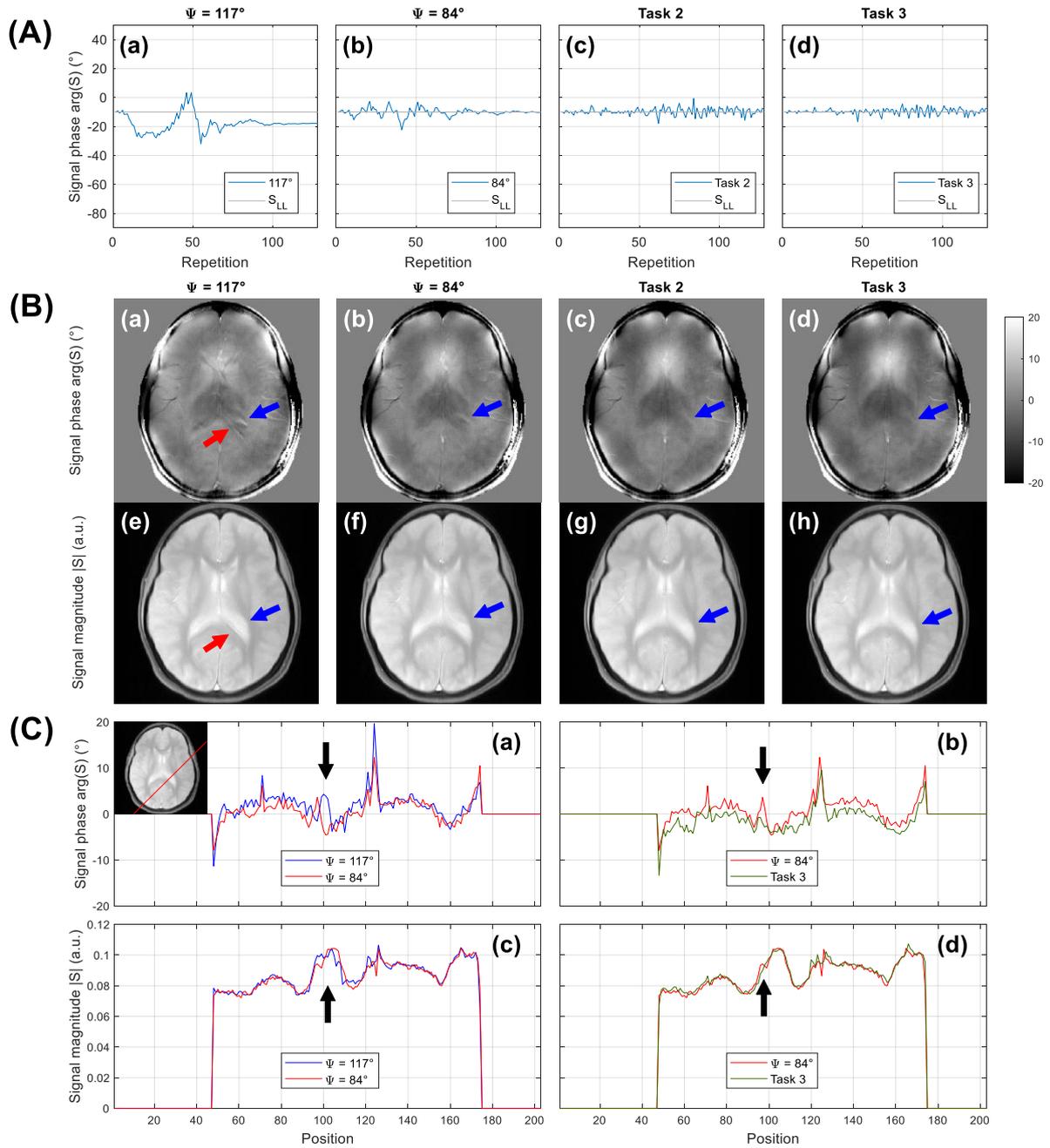

**Figure 9:** Unencoded simulation **(A)** of the FLASH signal using α = 19.5°, TE = 3.2 ms, TR = 6 ms for typical rf increments of 117° **(a)** and 84° **(b)** and for task 2 **(c)** and task 3 **(d)** are shown. By using a 2-shot sequence a resolution of 256 is achieved using the identical rf train two times, leading to a high-resolution phase image arg(S) **(Ba-d)** and magnitude image |S| **(Be-h)** of each FLASH sequence. The amount of the signal phase artefacts **(A)** corresponds to the artefacts observed in the phase images **(B)**. An ideal Look-Locker experiment should have a constant phase in the unencoded case, thus the smoothest phase image. The phase image of Ψ=117° shows artifacts in both unencoded signal phase **(Aa)** and image phase **(Ba)**, which is improved by Ψ=84° **(Ab,Bb)**, and further improved by our optimized sequence **(Ac-d,Bc-d)**. Red and blue arrows indicate clearly visible artifacts in the phase and magnitude images, which can be eliminated by the optimization tasks. **(C)** line profiles for signal phase **(a,b)** and signal magnitude **(c,d)** are chosen as shown in the inlay in **(Ca)**. A clear improvement can be achieved by using a quadratic increment of 84° instead of 117°. The artefact around position 103 can be eliminated in the magnitude signal, as well as in the phase **(Ca,Cc)**. Task 3 even outperforms a quadratic phase increment of 84° where remaining artefacts (position 96) vanish **(Cb,Cd)**.

## List of Table

**Table 1:** **(A)** RMSE with regard to the ideally spoiled signal $S_{LL}$, calculated for measured (Meas) and simulated (Sim) signal magnitude (left value) and phase (right value) for $\Psi = 117°,\ 84°$ and $50°$, and the optimized rf trains of task 1-3. For the RMSE of the PSF just the absolute signal is considered for calculation. The underlying signals are shown in Figure 1 and Figure 2. The corresponding normalized PSF is displayed in Figure 3. **(B)** RMSE w.r.t. $S_{LL}$ target system parameter changes (T2, T1, MI and rB1) for typical rf cycling of $117°$, $84°$ and $50°$, and the optimization tasks 1-3 corresponding to the signals shown in Figure 5 and 6. The value in the left column is the error of the signal magnitude, the right column is the error of the signal phase.

| (A) | $\Psi = 117°$ | | $\Psi = 84°$ | | $\Psi = 50°$ | | task 1 | | task 2 | | task 3 | |
|---|---|---|---|---|---|---|---|---|---|---|---|---|
| | Magnitude | Phase | Magnitude | Phase | Magnitude | Phase | Magnitude | Phase | Magnitude | Phase | Magnitude | Phase |
| Meas | 12.0·1e-3 | 2.23° | 5.4·1e-3 | 0.68° | 7.0·1e-3 | 0.93° | 2.6·1e-3 | 2.62° | 5.8·1e-3 | 0.35° | 3.0·1e-3 | 0.53° |
| Sim | 11.5·1e-3 | 2.02° | 5.5·1e-3 | 0.28° | 6.6·1e-3 | 0.49° | 2.2·1e-3 | 1.99° | 5.2·1e-3 | 0.21° | 1·1e-3 | 0.12° |
| PSF Meas | 15.0·1e-4 | | 6.3·1e-4 | | 7.6·1e-4 | | 14.0·1e-4 | | 5.7·1e-4 | | 4.8·1e-4 | |
| PSF Sim | 15.0·1e-4 | | 5.8·1e-4 | | 6.7·1e-4 | | 14.0·1e-4 | | 5.4·1e-4 | | 1.6·1e-4 | |
| B) | $\Psi = 117°$ | | $\Psi = 84°$ | | $\Psi = 50°$ | | task 1 | | task 2 | | task 3 | |
| | Magnitude | Phase | Magnitude | Phase | Magnitude | Phase | Magnitude | Phase | Magnitude | Phase | Magnitude | Phase |
| T2 | 17.9·1e-4 | 2.13° | 9.2·1e-4 | 0.34° | 11.0·1e-4 | 0.68° | 6.3·1e-4 | 2.11° | 9.1·1e-4 | 0.41° | 4.8·1e-4 | 0.33° |
| T1 | 16.2·1e-4 | 2.06° | 7.9·1e-4 | 0.31° | 9.2·1e-4 | 0.50° | 3.9·1e-4 | 2.05° | 7.2·1e-4 | 0.22° | 1.4·1e-4 | 0.14° |
| MI | 15.0·1e-4 | 13.03° | 6.9·1e-4 | 2.66° | 8.2·1e-4 | 4.04° | 9.2·1e-4 | 12.03° | 5.9·1e-4 | 8.14° | 2.3·1e-4 | 5.57° |
| rB1 | 58.7·1e-4 | 6.17° | 28.2·1e-4 | 0.89° | 31.6·1e-4 | 1.65° | 45.6·1e-4 | 5.98° | 27.2·1e-4 | 0.76° | 21.4·1e-4 | 0.58° |
| Total | 27.0·1e-4 | 5.85° | 13.1·1e-4 | 1.05° | 15.0·1e-4 | 1.72° | 16.3·1e-4 | 5.54° | 12.4·1e-4 | 2.38° | 7.5·1e-4 | 1.66° |

# References

1. Haase A, Frahm J, Matthaei D, Hanicke W, Merboldt KD. FLASH imaging. Rapid NMR imaging using low flip-angle pulses. *Journal of Magnetic Resonance (1969)*. 1986;67(2):258-266. doi:10.1016/0022-2364(86)90433-6

2. Crawley AP, Wood ML, Henkelman RM. Elimination of transverse coherences in FLASH MRI. *Magnetic Resonance in Med*. 1988;8(3):248-260. doi:10.1002/mrm.1910080303

3. Zur Y, Wood ML, Neuringer LJ. Spoiling of transverse magnetization in steady-state sequences. *Magnetic Resonance in Med*. 1991;21(2):251-263. doi:10.1002/mrm.1910210210

4. Epstein FH, Mugler JP, Brookeman JR. Spoiling of transverse magnetization in gradient-echo (GRE) imaging during the approach to steady state. *Magn Reson Med*. 1996;35(2):237-245. doi:10.1002/mrm.1910350216

5. Denolin V, Azizieh C, Metens T. New insights into the mechanisms of signal formation in RF-spoiled gradient echo sequences. *Magn Reson Med*. 2005;54(4):937-954. doi:10.1002/mrm.20652

6. Mugler JP, Brookeman JR. Rapid three-dimensional T1-weighted MR imaging with the MP-RAGE sequence. *Magnetic Resonance Imaging*. 1991;1(5):561-567. doi:10.1002/jmri.1880010509

7. Zaiss M, Ehses P, Scheffler K. Snapshot-CEST: Optimizing spiral-centric-reordered gradient echo acquisition for fast and robust 3D CEST MRI at 9.4 T. *NMR in Biomedicine*. 2018;31(4):e3879. doi:10.1002/nbm.3879

8. Krishnamoorthy G, Nanga RPR, Bagga P, Hariharan H, Reddy R. High quality three-dimensional gagCEST imaging of in vivo human knee cartilage at 7 Tesla. *Magnetic Resonance in Med*. 2017;77(5):1866-1873. doi:10.1002/mrm.26265

9. Jones RA, Southon TE. A magnetization transfer preparation scheme for snapshot FLASH imaging. *Magnetic Resonance in Med*. 1991;19(2):483-488. doi:10.1002/mrm.1910190242

10. Dai Z, Ji J, Xiao G, et al. Magnetization Transfer Prepared Gradient Echo MRI for CEST Imaging. Aoki I, ed. *PLoS ONE*. 2014;9(11):e112219. doi:10.1371/journal.pone.0112219

11. Look DC, Locker DR. Time Saving in Measurement of NMR and EPR Relaxation Times. *Review of Scientific Instruments*. 1970;41(2):250-251. doi:10.1063/1.1684482

12. Ganter C. Analytical solution to the transient phase of steady-state free precession sequences. *Magnetic Resonance in Med*. 2009;62(1):149-164. doi:10.1002/mrm.21968

13. Mugler JP, Epstein FH, Brookeman JR. Shaping the Signal Response during the Approach to Steady State in Three-Dimensional Magnetization-Prepared Rapid Gradient-Echo Imaging Using Variable Flip Angles. *Magnetic Resonance in Med*. 1992;28(2):165-185. doi:10.1002/mrm.1910280202

14. Stehling MK. Improved signal in "snapshot" FLASH by variable flip angles. *Magnetic Resonance Imaging*. 1992;10(1):165-167. doi:https://doi.org/10.1016/0730-725X(92)90387-F

15. Loktyushin A, Herz K, Dang N, et al. MRzero - Automated discovery of MRI sequences using supervised learning. *Magnetic Resonance in Med*. 2021;86(2):709-724. doi:10.1002/mrm.28727

16. Endres J, Weinmüller S, Dang HN, Zaiss M. Phase distribution graphs for fast, differentiable, and spatially encoded Bloch simulations of arbitrary MRI sequences. *Magnetic Resonance in Medicine*. Published online 2023. doi:10.1002/mrm.30055

17. Yarnykh VL. Optimal radiofrequency and gradient spoiling for improved accuracy of $T_1$ and $B_1$ measurements using fast steady-state techniques. *Magnetic Resonance in Med*. 2010;63(6):1610-1626. doi:10.1002/mrm.22394

18. Weigel M. Extended phase graphs: Dephasing, RF pulses, and echoes - pure and simple. *Magnetic Resonance Imaging*. 2015;41(2):266-295. doi:10.1002/jmri.24619

19. Hennig J. Echoes—how to generate, recognize, use or avoid them in MR-imaging sequences. Part I: Fundamental and not so fundamental properties of spin echoes. *Concepts Magn Reson*. 1991;3(3):125-143. doi:10.1002/cmr.1820030302

20. Hennig J. Echoes—how to generate, recognize, use or avoid them in MR-imaging sequences. Part II: Echoes in Imaing Sequences. *Concepts Magn Reson*. 1991;3(3):179-192. doi:10.1002/cmr.1820030302

21. Hennig J. Multiecho imaging sequences with low refocusing flip angles. *Journal of Magnetic Resonance (1969)*. 1988;78(3):397-407. doi:10.1016/0022-2364(88)90128-X

22. Malik SJ, Padormo F, Price AN, Hajnal JV. Spatially resolved extended phase graphs: Modeling and design of multipulse sequences with parallel transmission. *Magnetic Resonance in Med*. 2012;68(5):1481-1494. doi:10.1002/mrm.24153

23. Paszke A, Gross S, Massa F, et al. PyTorch: An Imperative Style, High-Performance Deep Learning Library. In: *Advances in Neural Information Processing Systems 32*. Curran Associates, Inc.; 2019:8024-8035. http://papers.neurips.cc/paper/9015-pytorch-an-imperative-style-high-performance-deep-learning-library.pdf

24. MR-Physics-With-Pulseq. Accessed February 27, 2024. https://github.com/pulseq/MR-Physics-with-Pulseq

25. Leupold J, Hennig J, Scheffler K. Moment and direction of the spoiler gradient for effective artifact suppression in RF-spoiled gradient echo imaging. *Magnetic Resonance in Med*. 2008;60(1):119-127. doi:10.1002/mrm.21614

26. Scheffler K. A pictorial description of steady-states in rapid magnetic resonance imaging. *Concepts Magn Reson*. 1999;11(5):291-304. doi:10.1002/(SICI)1099-0534(1999)11:5<291::AID-CMR2>3.0.CO;2-J

27. Preibisch C, Deichmann R. Influence of RF spoiling on the stability and accuracy of $T_1$ mapping based on spoiled FLASH with varying flip angles. *Magnetic Resonance in Med*. 2009;61(1):125-135. doi:10.1002/mrm.21776

28. Kingma DP, Ba J. Adam: A Method for Stochastic Optimization. Published online 2014. doi:10.48550/ARXIV.1412.6980

29. Layton KJ, Kroboth S, Jia F, et al. Pulseq: A rapid and hardware-independent pulse sequence prototyping framework: Rapid Hardware-Independent Pulse Sequence Prototyping. *Magn Reson Med*. 2017;77(4):1544-1552. doi:10.1002/mrm.26235

30. Ravi K, Geethanath S, Vaughan J. PyPulseq: A Python Package for MRI Pulse Sequence Design. *JOSS*. 2019;4(42):1725. doi:10.21105/joss.01725

31. Ravi KS, Potdar S, Poojar P, et al. Pulseq-Graphical Programming Interface: Open source visual environment for prototyping pulse sequences and integrated magnetic resonance imaging algorithm development. *Magnetic Resonance Imaging*. 2018;52:9-15. doi:10.1016/j.mri.2018.03.008

32. BrainWeb: Simulated Brain Database. Accessed January 1, 2023. https://brainweb.bic.mni.mcgill.ca/brainweb/

33. Bojorquez JZ, Bricq S, Acquitter C, Brunotte F, Walker PM, Lalande A. What are normal relaxation times of tissues at 3 T? *Magnetic Resonance Imaging*. 2017;35:69-80. doi:10.1016/j.mri.2016.08.021

34. Peters AM, Brookes MJ, Hoogenraad FG, et al. T2* measurements in human brain at 1.5, 3 and 7 T. *Magnetic Resonance Imaging*. 2007;25(6):748-753. doi:10.1016/j.mri.2007.02.014

35. Le Bihan D, Mangin J, Poupon C, et al. Diffusion tensor imaging: Concepts and applications. *Magnetic Resonance Imaging*. 2001;13(4):534-546. doi:10.1002/jmri.1076

36. Terpstra M, Maspero M, Lagendijk J, van den Berg CAT. Rethinking complex image reconstruction: ⊥-loss for improved complex image reconstruction with deep learning. In: ; 2021. https://cds.ismrm.org/protected/21MPresentations/abstracts/1751.html

37. Budde J, Shajan G, Scheffler K, Pohmann R. Ultra-high resolution imaging of the human brain using acquisition-weighted imaging at 9.4T. *NeuroImage*. 2014;86:592-598. doi:10.1016/j.neuroimage.2013.08.013

38. Pohmann R, Von Kienlin M. Accurate phosphorus metabolite images of the human heart by 3D acquisition-weighted CSI. *Magnetic Resonance in Med*. 2001;45(5):817-826. doi:10.1002/mrm.1110

39. Sobol WT, Gauntt DM. On the stationary states in gradient echo imaging. *Magnetic Resonance Imaging*. 1996;6(2):384-398. doi:10.1002/jmri.1880060220

40. Kir A, McMillan A. Optimized inversion-prepared gradient echo imaging. *Magnetic Resonance Imaging*. 2012;36(3):748-755. doi:10.1002/jmri.23669

41. Priatna A, Paschal CB. Variablemangle uniform signal excitation (vuse) for threel dimensional time-of-flight MR angiography. *Magnetic Resonance Imaging*. 1995;5(4):421-427. doi:10.1002/jmri.1880050409

42. Stöcker T, Shah NJ. MP-SAGE: A new MP-RAGE sequence with enhanced SNR and CNR for brain imaging utilizing square-spiral phase encoding and variable flip angles. *Magn Reson Med*. 2006;56(4):824-834. doi:10.1002/mrm.21011

43. Li X, Han ET, Busse RF, Majumdar S. In vivo $T_{1\rho}$ mapping in cartilage using 3D magnetization-prepared angle-modulated partitioned $k$-space spoiled gradient echo snapshots (3D MAPSS). *Magnetic Resonance in Med*. 2008;59(2):298-307. doi:10.1002/mrm.21414

44. Worters PW, Hargreaves BA. Balanced SSFP transient imaging using variable flip angles for a predefined signal profile: Variable Flip Angles bSSFP Transient Imaging. *Magn Reson Med*. 2010;64(5):1404-1412. doi:10.1002/mrm.22541

45. Tran-Gia J, Wech T, Bley T, Köstler H. Model-Based Acceleration of Look-Locker T1 Mapping. Langowski J, ed. *PLoS ONE*. 2015;10(4):e0122611. doi:10.1371/journal.pone.0122611

46. Maier O, Schoormans J, Schloegl M, et al. Rapid $T_1$ quantification from high resolution 3D data with model-based reconstruction. *Magnetic Resonance in Med*. 2019;81(3):2072-2089. doi:10.1002/mrm.27502

47. Wang X, Scholand N, Tan Z, et al. Model-Based Reconstruction for Joint Estimation of $T_{1}$, $R_{2}^{*}$ and $B_{0}$ Field Maps Using Single-Shot Inversion-Recovery Multi-Echo Radial FLASH. Published online 2024. doi:10.48550/ARXIV.2402.05366

48. Scholand N, Wang X, Roeloffs V, Rosenzweig S, Uecker M. Quantitative Magnetic Resonance Imaging by Nonlinear Inversion of the Bloch Equations. Published online September 16, 2022. Accessed January 12, 2023. http://arxiv.org/abs/2209.08027

49. Dang HN, Endres J, Weinmüller S, et al. MR-zero meets RARE MRI: Joint optimization of refocusing flip angles and neural networks to minimize $T_2$-induced blurring in spin echo sequences. *Magnetic Resonance in Med*. 2023;90(4):1345-1362. doi:10.1002/mrm.29710

50. Hess AT, Robson MD. Hexagonal gradient scheme with RF spoiling improves spoiling performance for high-flip-angle fast gradient echo imaging. *Magnetic Resonance in Med*. 2017;77(3):1231-1237. doi:10.1002/mrm.26213

51. Chow K, Flewitt JA, Green JD, Pagano JJ, Friedrich MG, Thompson RB. Saturation recovery single-shot acquisition (SASHA) for myocardial $T_1$ mapping. *Magnetic Resonance in Med*. 2014;71(6):2082-2095. doi:10.1002/mrm.24878

52. Liu C, Li W, Tong KA, Yeom KW, Kuzminski S. Susceptibility-weighted imaging and quantitative susceptibility mapping in the brain: Brain Susceptibility Imaging and Mapping. *J Magn Reson Imaging*. 2015;42(1):23-41. doi:10.1002/jmri.24768

# Supporting Information

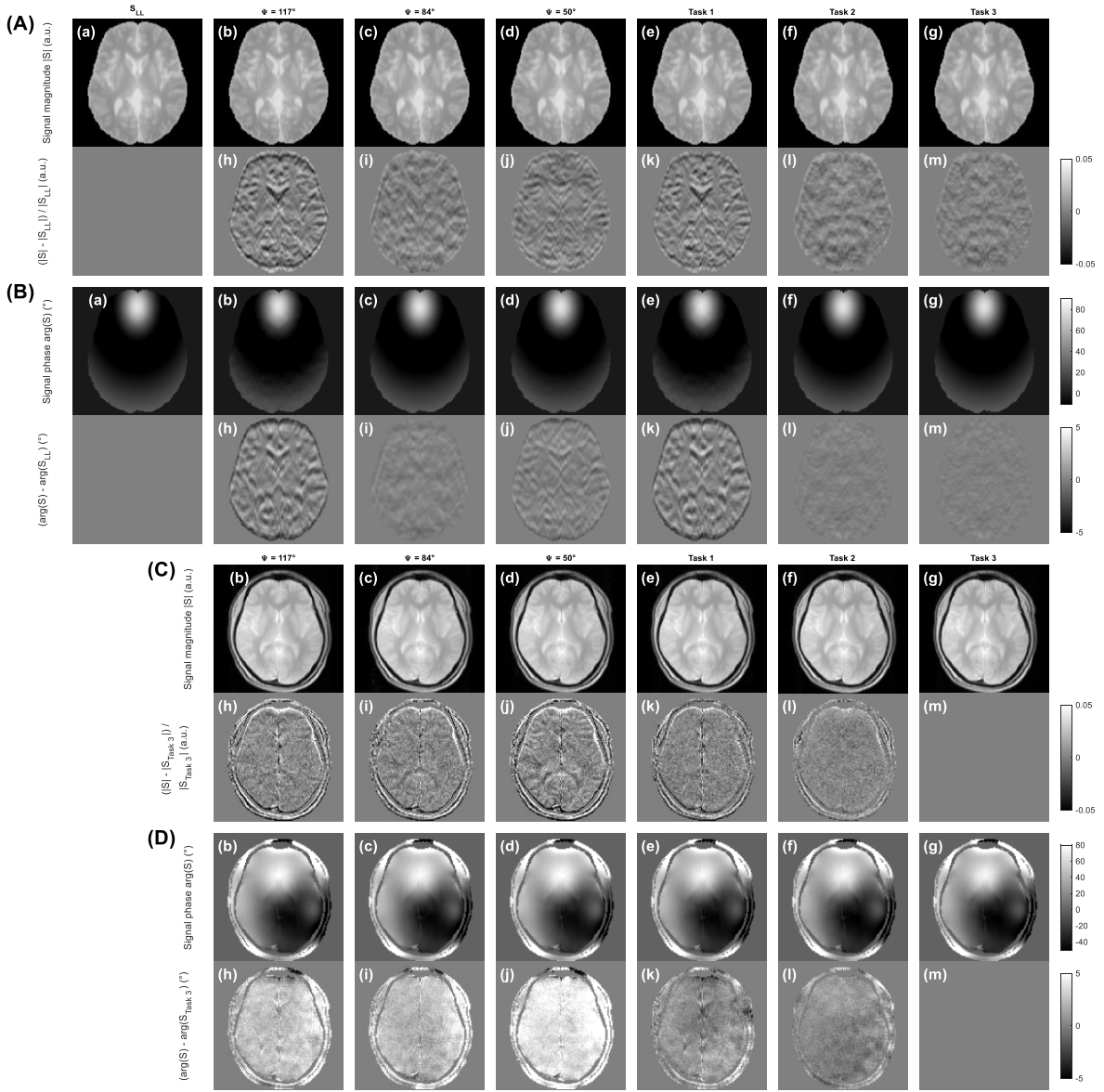

**Supporting Information Figure S1:** Simulation **(A, B)** and measurement **(C, D)** of centric encoded FLASH readouts using α = 10°, TE = 5.0 ms, TR = 10.0 ms for typical rf increments of 117° **(b)**, 84° **(c)** and 50° **(d)** and optimization task 1 **(e)**, task 2 **(f)** and task 3 **(g)**. In simulation the ideal spoiled Look-Locker image $S_{LL}$ was additional calculated. The phantom for simulation was built from data provided by the BrainWeb database. There, differences of signal magnitude S **(Ah-m)** and phase arg(S) **(Bh-m)** are calculated for each image to the Look-Locker target **(a)**. For the magnitude images **(Cb-g)** and phase images **(Db-g)** task 3 is used as reference **(g)** in measurement. Artefacts in the phase image can be seen for all standard rf increments, especially for 117°. Task 3 can reduce these artefacts coming from the rf cycling and generate the smoothest phase image. In vivo fewer clear artefacts are visible compared to Figure 7 with higher excitation flip angle a longer TE and TR. Susceptibility effects are in the same order of magnitude. The seen offset in the phase differences **(Dg-Di)** can be explained by the shifting B0 field of the used scanner.

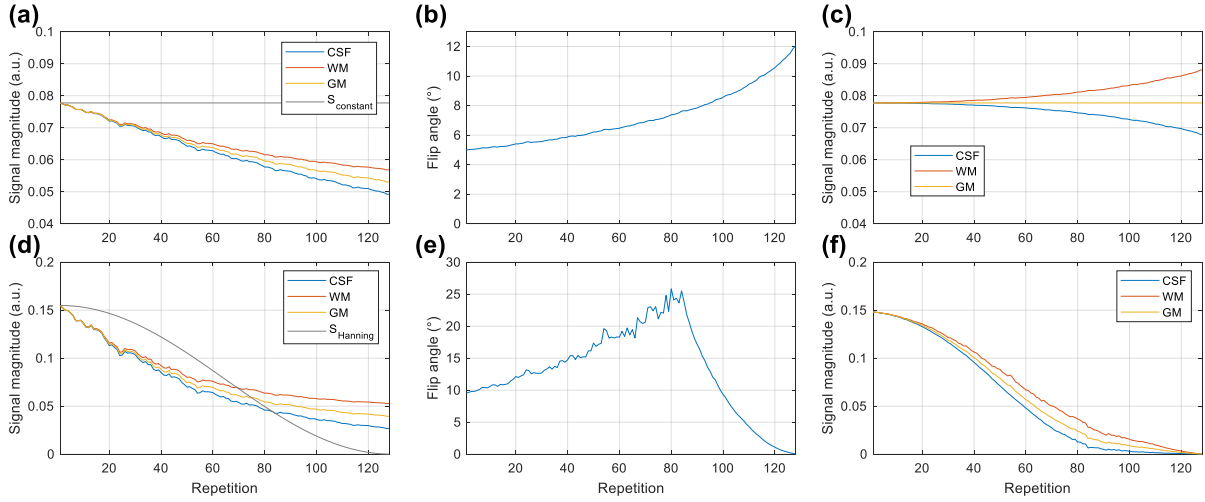

**Supporting Information Figure S2:** Signal evolution of a voxel with different T1 values (CSF – 4.0 s, WM – 0.83 s and GM – 1.48 s) are shown in **(a,d)**. Additionally, in grey the constant signal target **(a)** and Hanning function target (h(n) = $\frac{\beta}{2}\left(1 + \cos\frac{2\pi n \Delta x}{\alpha'}\right)$ with $\beta = 1, \alpha' = 2, \Delta x = 1/128$) **(d)** is plotted. For task 4 ($\vec{\alpha} \mid S_{const}$; $\Psi = 84°$) and task 5 ($\vec{\alpha} \mid S_{Hanning}$; $\Psi = 84°$) the final optimized flip angle trains are given in **(b,e)**. In **(c,f)** the signal magnitude evolution is visualize again for different T1 values. The sequence parameters in the shown optimization tasks are α = 5° for task 4 and α = 10° for task 5, a readout bandwidth of 500 Hz/pixel, TE = 3.2 ms and TR = 6 ms is used. No magnetization preparation is applied before the readout in this case. With flip angle optimization the target signal can be reached, at least for the in training used T1 value. Here, the training is performed with the median of a BrainWeb phantom, which has a GM-like T1 value. However, the optimized flip angle train clearly fails for another T1 values (WM and CSF). The solution is tissue and problem depended.

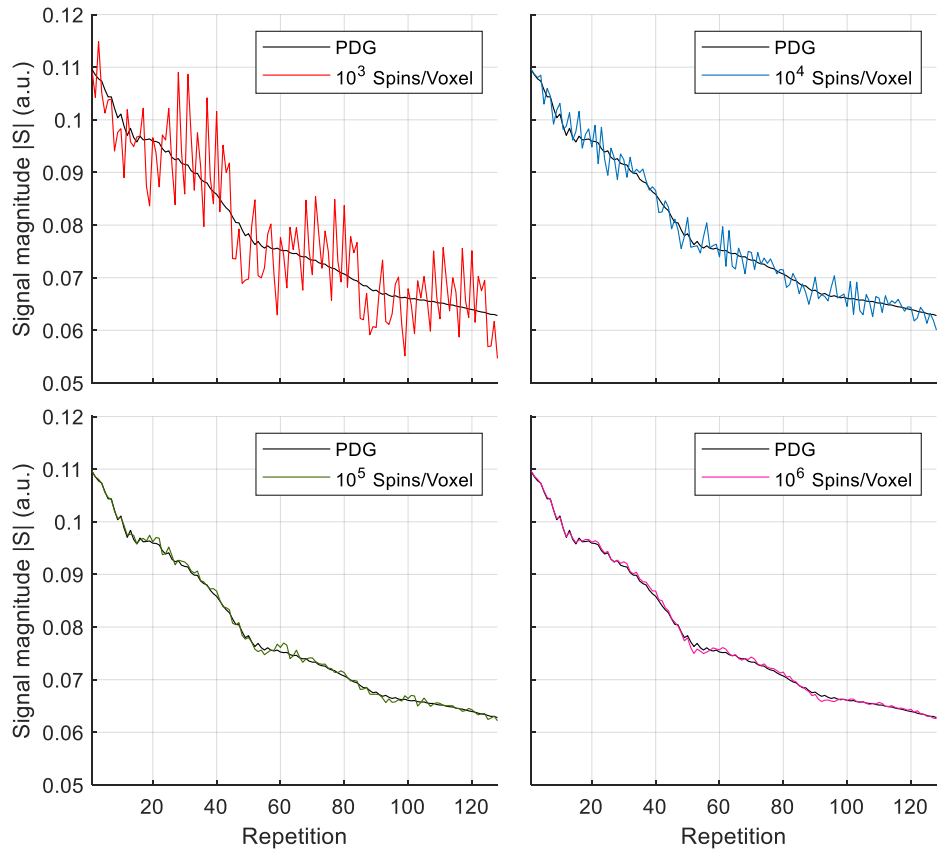

**Supporting Information Figure S3:** Comparison between spin simulation (colored line) using different number of spins per voxel and the same PDG simulation (black line).

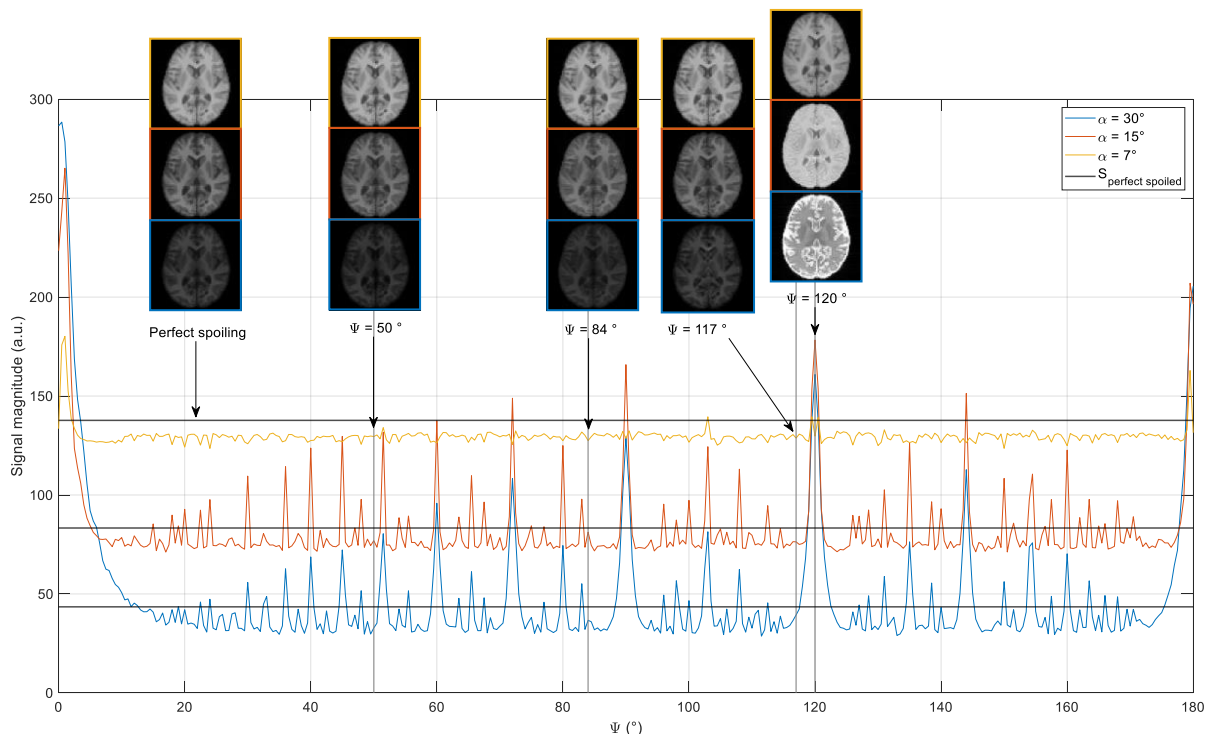

**Supporting Information Figure S4:** Steady-state signal magnitude of FLASH sequence (TR = 5.6 ms) are shown as function of the spoiling phase increment $\Psi$ for different excitation flip angles for a white matter pixel (T1 = 0.85 s, T2 = 0.076 s, D = 0.66 ·10$^{-3}$ mm$^2$/s). For typical rf phases the images are shown. The signal magnitude of a perfect spoiled sequence doesn't depend on the phase increment (grey color). The corresponding image is visualized as well. Strong deviations from a perfect spoiled steady state signal can be seen for higher excitation flip angles.

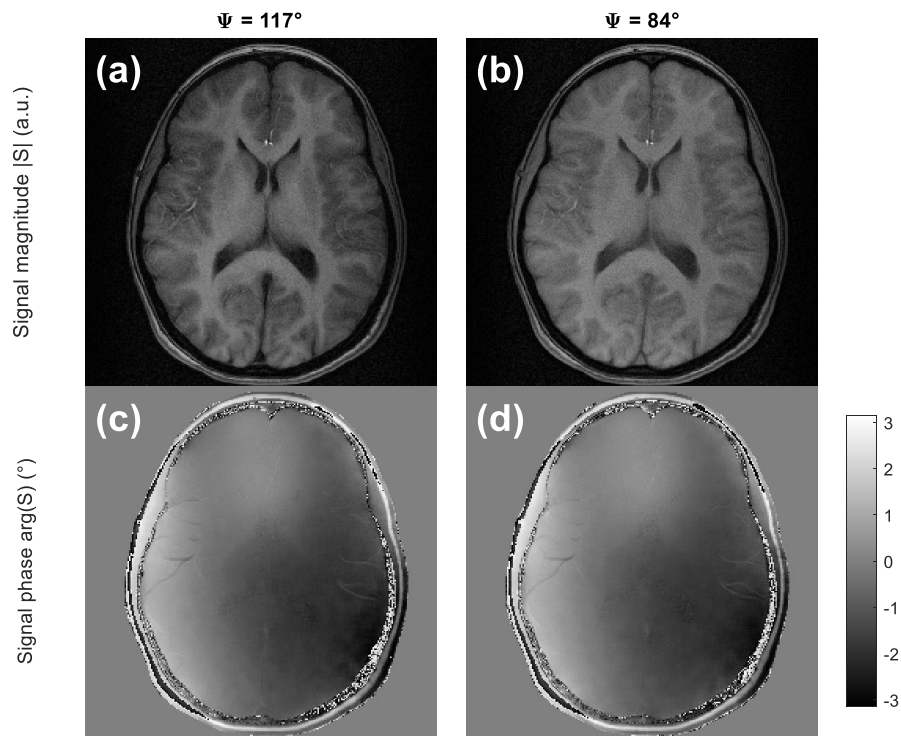

**Supporting Information Figure S5:** Standard clinical MRRAGE magnitude **(a,b)** and phase **(c,d)** images using typical quadratic rf increments of $\Psi = 117°$ and $\Psi = 84°$. Sequence parameter of linear-reordered MPRAGE: inversion time: 0.9 s; resolution: 256x256; FOV: 200x200 mm$^2$; TE: 4.2 ms; echo distance: 8.2 ms; TR: 1.95 s bandwidth: 280 Hz/pixel; average: 2; shots: 2; dummy shots: 3. No significant difference can be seen in magnitude and phase image produced by different rf increments, just a slightly different contrast is visible in the magnitude.

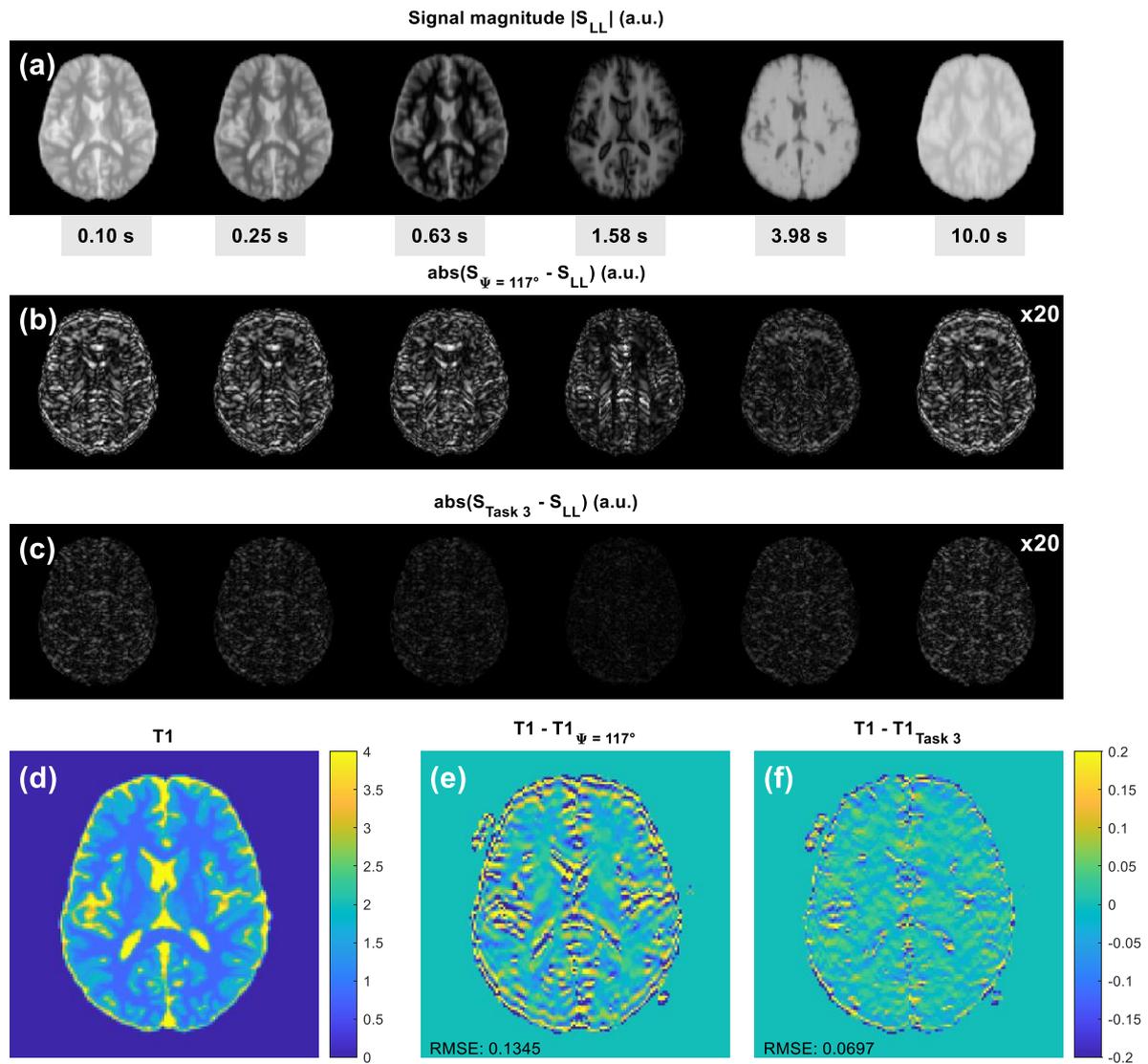

**Supporting Information Figure S6:** The magnitude images for different inversion times given in the gray box are shown in **(a)**. In this simulation, the ideal Look-Locker decay is considered. A centric encoded FLASH readout was used with parameters α = 19.5°, readout bandwidth of 500 Hz/pixel, TE = 3.2 ms and TR = 6 ms. Before each inversion we assume to be in a completely relaxed thermal equilibrium state. Subfigures **(b)** and **(c)** display the absolute difference between **(a)** and a full simulation that uses all necessary signal distributions for a quadratic rf increment of $\Psi = 117°$, as well as the optimized sequence from task 3. The optimized rf train reduces clearly the error of the magnitude images. Then we quantify T1 for these both datasets using a Look-Locker model for reconstruction. The T1 map of the BrainWeb phantom is presented in **(d)**. The Look-Locker model leads to T1 map differences for the rf cycling of 117° **(e)**, which are mitigated when using the data acquired with the optimized rf train of task 3 **(f)**, which is also indicated by a lower RMSE.